# Differential Perspectives: Epistemic Disconnects Surrounding the US Census Bureau's Use of Differential Privacy

danah boyd, Microsoft Research and Data & Society
Jayshree Sarathy, Harvard University

**Abstract:** When the U.S. Census Bureau announced its intention to modernize its disclosure avoidance procedures for the 2020 Census, it sparked a controversy that is still underway. The move to differential privacy introduced technical and procedural uncertainties, leaving stakeholders unable to evaluate the quality of the data. More importantly, this transformation exposed the statistical illusions and limitations of census data, weakening stakeholders' trust in the data and in the Census Bureau itself. This essay examines the epistemic currents of this controversy. Drawing on theories from Science and Technology Studies (STS) and ethnographic fieldwork, we analyze the current controversy over differential privacy as a battle over uncertainty, trust, and legitimacy of the Census. We argue that rebuilding trust will require more than technical repairs or improved communication; it will require reconstructing what we identify as a 'statistical imaginary.'

## 1. Introduction

Every decade since 1790, the United States has embarked on a grand effort to count the whole population in order to apportion political representatives and, more recently, allocate federal funding and uphold a range of government policies (M. J. Anderson, 2015). The census data that are produced by the federal government serve as democracy's data infrastructure (Bouk & boyd, 2021). Each decadal count is rife with its own controversies (e.g., Alonso et al., 1987; National Research Council (U.S.) et al., 2004), in no small part because the data are so politically and scientifically important. The 2020 Census is no different.

One of the controversies surrounding the 2020 Census concerns the Census Bureau's decision to implement a new disclosure avoidance system based on the mathematical framework of differential privacy. This controversy is ongoing. It is also the subject of this special issue. The other articles in this special issue focus on the technical merits and limitations of this system, revealing disagreements about how both data quality and the system itself can and should be interpreted. The goal of this essay is to put these disagreements – and those that exist outside of this special issue – into perspective using the theoretical tools of Science and Technology Studies (STS). While the current 'fight' concerning differential privacy is being debated in legal, technical, and procedural terms, our perspective is that the root of the controversy is epistemic in nature.

Epistemology is the study of knowledge, how people know what they know. Communities of practice – including both scientific and technical communities – form around shared epistemic worldviews (Lave & Wenger, 1991). Controversies often call into question the boundaries of a given community and worldview (Gieryn, 1983), revealing fractures in the community's epistemic commitments. A cornerstone of the field of science studies is the examination of scientific controversy (e.g., Engelhardt & Caplan, 1987; Machamer et al., 2000). Notably, scholars from this discipline often ask what is required for a scientific controversy to end. Controversies may achieve closure through loss of interest, consensus, sound argument, negotiation, or force (Beauchamp, 1987; McMullin, 1987). Through this process, epistemic fractures are navigated, and boundaries are formed.

Emergent technical systems introduce another layer to this dynamic. Technological systems are socially constructed and controversies often emerge when different stakeholders interpret a given technology in different lights, a process that technology scholars refer to as 'interpretive flexibility' (Bijker et al., 2012). Technical systems are stabilized when the controversy surrounding the technology achieves closure. Yet, as Wieber Bijker and Trevor Pinch note, closure does not mean that the problem is 'solved.' Rather, closure often emerges when the perceived problem disappears ('rhetorical closure') or when the problem itself is redefined (Pinch & Bijker, 1984).

Scientific and technical controversies reveal epistemic, ideological, and political commitments within communities of practice, but politically motivated actors outside of a given community of practice can also leverage epistemic fractures to pursue other agendas (Jasanoff, 2010; Proctor & Schiebinger, 2008). For example, such actors have leveraged epistemic fractures in government to stall legislation around tobacco, climate change, and nuclear power.

As our democracy's data infrastructure, the census is particularly vulnerable to this kind of political disruption; litigious fights and politicization have grown over the last century (Bryant & Dunn, 1995; Choldin, 1994). With the emergence of a new technical system for disclosure avoidance, and the continuous disputes around what constitutes a legitimate count during the height of a pandemic, the 2020 Census has provided the perfect storm for a controversy. The Census Bureau's disclosure avoidance system has become a lightning rod, prompting outrage, confusion, and lawsuits. Among the technical community engaged with the census – including data users, statisticians, and computer scientists – the disclosure avoidance system has sparked an epistemic reckoning about data quality. This system has also forced legal and policy experts to reevaluate what the work and responsibilities of a statistical agency are. Scientific and policy debates have spilled out into non-technical contexts, as various stakeholders seek to convince other members of the ecosystem of their epistemic perspective and enroll them into assorted political and advocacy efforts, some of which have resulted in lawsuits directed at the Census Bureau. Among many census stakeholders, public discussions about this system and the choices the bureau made have shattered a 'statistical imaginary,' a concept we unpack later in this essay, upon which the contemporary Census Bureau's legitimacy resides.

The controversy around differential privacy and the 2020 Census is by no means settled. As of this writing, legal cases are both ongoing and emerging.[1] This special issue reflects some of the ongoing technical debates. In other spheres, policymakers and advocates are discussing revisions to Title 13 of the U.S. Code, which regulates the Census Bureau's privacy and confidentiality requirements (and distinguishes between those two concepts).[2] Data users in academia, tribal, state, and local governments, the professional redistricting community, and NGOs continue to convene to discuss different issues related to the Census Bureau's disclosure avoidance system, looking for ways to challenge the bureau's plans. Most of these conversations are not open to the public, but those who attend repeatedly describe them as contentious. The Census Bureau continues to seek stakeholder feedback through commentary processes and in conjunction with their Federal Advisory Committees and the Committee on National Statistics.[3] Feedback given to the bureau repeatedly questions the validity and legitimacy of the new disclosure avoidance efforts. In other words, even though the Census Bureau has already published the 2020 redistricting data using a system that relied on differential privacy, closure to this controversy is not on the immediate horizon.

---

[1] *Alabama v. U.S. Dep't of Commerce* was dismissed in September of 2021, but the same lawyers and experts have subsequently filed a new case - *Fair Lines America Foundation v. U.S. Dep't of Commerce.* While this lawsuit is purportedly about obtaining documents pertaining to the counting of group quarters, differential privacy has emerged as a central topic. Meanwhile, redistricting advocates are debating whether to bring additional lawsuits.

The papers and essays in this special issue – including this one – seek to illustrate and shape the contours of this controversy. We are offering this essay to help make visible the epistemic fractures that surround differential privacy and the 2020 Census. Following an overview of the controversy, we examine two sites of conflict that illustrate how epistemic disagreements structure the contestations. We then turn to reflect on what we see as an epistemic fracture to a statistical imaginary, a fracture that we argue requires re-thinking how the statistical work of the U.S. Census Bureau is understood in order to repair the legitimacy of the census.

*A Note on Method*

The arguments in this paper are themselves derived from different epistemic vantage points. One author has conducted four years of ethnographic fieldwork among census stakeholders (including 47 interviews related to this topic); the other author has been an active contributor to a broader, open-source differential privacy initiative called OpenDP.[4] This paper draws on our experiential and ethnographic data, as well as on material from over a dozen public events, on social media and blogs, and in court filings and news media. Throughout this paper, we describe postures and reactions held by government officials and external stakeholders (including representatives from civil rights groups and local/state governments, as well as academics, redistricters, and other data users); these sentiments were communicated to at least one of us directly during interviews or private conversations, or in small group meetings that we attended. To ensure that we depicted these attitudes appropriately, drafts of this essay were shared with 54 stakeholders representing different perspectives detailed in this paper as a form of 'member checking' our arguments; we incorporated their feedback into our edits. When we depict a point of view held by a particular group of actors, we are drawing from a minimum of three examples with few (if any) counterexamples from that group.

---

[2] Most proposed revisions to Title 13 are being discussed privately, but some census experts are publicly calling for changes. See, for example: https://magazine.amstat.org/blog/2022/02/01/data-infrastructure-census/
[3] The Census Bureau's stakeholder engagement work can be found here: https://www.census.gov/programs-surveys/decennial-census/decade/2020/planning-management/process/disclosure-avoidance.html
[4] https://opendp.org/

This paper attempts to grapple with divergent, incompatible, and contradictory epistemic frames to examine the contestation at play, including the ones that have shaped our thinking and that of our informants.

## 2. Background to Privacy and Confidentiality

In 2003, a team of computer scientists proved that given enough statistical information about a database, one can reconstruct the entire database, thereby revealing potentially sensitive information. They detailed how statistical tabulations 'leak' privacy and how, given enough leakage, the 'reconstruction algorithm' they provide can perfectly recreate the individual data records (Dinur & Nissim, 2003). This paper initiated a flurry of research examining database vulnerabilities and prompted four cryptographers to devise a remedy (Dwork et al., 2006). The remedy – later named 'differential privacy' – is a mathematical framework for measuring and controlling the injection of noise into statistical releases to ensure statistical confidentiality and enable data usability. Differential privacy is not a system, but rather a mathematical definition of privacy that system-builders can use to evaluate their privacy-preserving mechanisms (Dwork et al., 2011; Wood et al., 2018). For every type of statistical information (e.g., mean, median, sum) a system publishes about the underlying data, a differentially private algorithm can be used to control the privacy-loss within that release. The noise across these releases can be calibrated such that the total privacy-loss stays within a designated 'privacy-loss budget.'

The more statistics that a system publishes, the greater the risk to statistical confidentiality. The U.S. Census Bureau publishes billions of statistics from its data. Yet, for procedural, legal, moral, and historical reasons, statistical confidentiality is a deeply held commitment among government workers. Initially, from 1790-1840, US census records were not kept confidential; after the data was collected, U.S. marshals were required to post the results in the town square for all to see (U.S. Census Bureau, 2019). The rationale for public scrutiny was quite sensible. By making census data visible for inspection, members of the community could ensure that no mistakes were made, and no people were missed. But the government soon realized that the very act of posting this data left many people wary of answering the census taker's questions. For a democracy "to ensure the public trust necessary to carry out reliable statistical inquiries," the government needed to intentionally, and continually, separate the census from surveillance in the minds of the public (Starr, 1987, p. 13). Thus, from 1850 onwards, individual census records became mostly inaccessible to the public, first by norm and then by law.

The Reapportionment Act of 1929 (2 U.S.C. § 2a) attempted to define statistical confidentiality by explicitly stating that census data could only be used for statistical purposes: "No publication shall be made by the Census Office whereby the data furnished by any particular establishment or individual can be identified." Later, when Title 13 of the U.S. Code was created in 1954 to regulate future censuses, the earlier confidentiality provision was codified into the new law, in part to resolve controversies over confidentiality during that era (Bryant & Dunn, 1995, p. 33). As a result of Title 13, all who work at the U.S. Census Bureau must swear a lifetime oath to uphold the confidentiality commitments. This oath serves as a mantra at the bureau, shaping every aspect of organizational life.

The bureau's moral commitment to statistical confidentiality has had a more sordid history. As historians William Seltzer and Margo Anderson (Seltzer & Anderson, 2000, 2007) uncovered at the turn of the 21st century, census data had been used in efforts to incarcerate Americans of Japanese ancestry in internment camps during World War II. Although this was not illegal at the time, as the Second War Powers Act (50a U.S.C. §§ 633-645) overturned the protections laid out in the 1929 Census Act, this breach of statistical confidentiality violated both longstanding norms as well as moral commitments. Data released after the 2000 census prompted a second reckoning over moral responsibilities after privacy advocates discovered that the Census Bureau had provided special tabulations of Arab-Americans to the Department of Homeland Security (El-Badry & Swanson, 2005; Electronic Privacy Information Center, 2004).

In his paper on "ethics, confidentiality, and data dissemination," Hermann Habermann, a former Deputy Director of the U.S. Census Bureau, reckoned with both the World War II history as well as the special tabulation (Habermann, 2006). Habermann defended the bureau's actions while urging future government agencies to grapple with the moral aspects of the work they do: "Government is more than bureaucracy, and individuals do exercise moral authority and must make moral judgments. Adherence to bureaucratic procedures may, in fact, be a denial of the basic responsibility of those who work in statistical agencies. The common good at any particular time may negate the particular good of the statistical agency and/or of particular groups" (Habermann, 2006, p. 614). While rejecting Habermann's defense of the bureau's decisions, Stephen Fienberg – a professor of statistics known for his work on privacy – also emphasized values over law: "Sharing data is a matter of ethics and the U. S. Census Bureau's data are a public good. So the need to provide greater access to Bureau data seems obvious to me. But I also see the possibility that some data can be misused, either by government officials or by others who access them. There is an ethical obligation not to aid and abet that abuse" (Fienberg, 2006). Both men recognized the entanglement of legal, moral, and procedural commitments, even as they viewed the ethics of individual decisions differently.

As the 1990 census director Barbara Bryant predicted (Bryant & Dunn, 1995), the combination of historical controversies as well as procedural, legal, and moral duties have shaped the Census Bureau's commitment to disclosure avoidance in the 21st century. Meanwhile, the vulnerabilities introduced by Dinur & Nissim's reconstruction attack (Dinur & Nissim, 2003) called into question the traditional disclosure avoidance procedures that the bureau had been using.

The practices used to achieve statistical confidentiality have evolved significantly since 1850, when only the most essential statistics were tabulated and published in large physical volumes. After computers made tabulation and data-sharing easier, the 1970 Census began performing 'table suppression,' intentionally choosing specific tables to *not* publish in order to ensure that the data remained confidential (McKenna, 2018). For some data products, rounding and 'top coding'[5] also became a routine practice for ensuring statistical confidentiality in addition to table suppression.

---

[5] 'Top coding' is a technique to reduce the visibility of outliers with large values by applying an upper bound to the data. Rather than reporting on single year of age for every individual, for example, the Census Bureau might denote the age of all who are older than 100 years of age as 100.

While table suppression, rounding, and top-coding provided one avenue for protecting census data, these techniques prevented the Census Bureau from publishing small-area statistics or data about less populous demographics. Census Bureau researchers began evaluating 'controlled perturbation'[6] approaches in the 1980s. For the 1990 Census, under pressure to publish 'block-level' data for the entire country, the Census Bureau altered the disclosure avoidance processes it used in the census by introducing noise-infusion procedures like 'swapping'[7], 'blank-and-impute'[8], and 'partially synthetic data'[9] (McKenna, 2018). All these noise-infusion procedures altered census data through procedural steps designed to minimize the visibility of outliers, often by 'smoothing' the data, but details about the application of these procedures could not be made public without weakening the confidentiality protections.[10] This meant that there was no way to assess how the noise might have biased statistical calculations.

Tracking advances in computer science, the U.S. Census Bureau decided to first experiment with differential privacy within a year of its development. In 2008, the economic directorate released OnTheMap, a tool that allowed users to visualize data from the longitudinal employer-household dynamics program (Machanavajjhala et al., 2008). John Abowd – an economics and statistics professor at Cornell (and at the time, a fellow at the Census Bureau) – led the confidentiality protection work for this effort. This tool was notable because it showed how differential privacy would allow the bureau to release previously unpublished data without risking the confidentiality of the underlying records.

The imminence of the 2010 Census made it impossible for bureau executives to consider implementing a new disclosure avoidance system that relied on differential privacy. Still, as multiple bureau executives told us, the 2010 Census leadership team was concerned that the published census data might be vulnerable. Rather than reverting to table suppression and reveal the vulnerability, however, the leadership team decided to use the same disclosure avoidance procedures in 2010 that had been used in 2000, relying heavily on swapping and public ignorance. After the 2010 Census was completed and all of the data were published, executives asked John Abowd to take leave from Cornell and become an Associate Director and the Chief Scientist at the Census Bureau. Upon arrival, he was assigned to lead a team to assess the vulnerability of the 2010 data to a reconstruction attack.[11] The findings from this internal analysis prompted the Data Stewardship Executive Policy Committee at the Census Bureau to begin investigating the possibility of modernizing the disclosure avoidance procedures for the 2020 Census to incorporate differential privacy.

---

[6] 'Controlled perturbation' references techniques used to systematically control the injection of noise into statistics (Cox et al., 1986). Differential privacy is a technique to mathematically measure such perturbations.

[7] In the decennial census, 'swapping' involves taking an outlier household from one geography and swapping it with a household in a nearby geography (Gatewood & Micarelli, 2001)

[8] 'Imputation' is a procedure for addressing missing data; the decennial census procedure for 'characteristic' or 'item' imputation involves taking the demographic characteristics from a nearby household and applying those to a household lacking information (Cantwell et al., 2004). 'Blank and impute' involves erasing the reported characteristics of an outlier household and then applying the imputation procedure to ensure the confidentiality of that outlier household.

[9] 'Partially synthetic data' are produced using more sophisticated imputation procedures designed for statistical validity and with an eye for disclosure avoidance (Reiter & Kinney, 2012).

[10] In other words, these noise-infusion procedures rely on 'security by obscurity,' which is heavily discouraged in the cryptography and security community.

[11] These findings have been reported on at conferences, webinars, and advisory meetings. Additional detail can be found in John Abowd's declaration in *Alabama v. Commerce* (US. District Court for the Middle District of Alabama Eastern Division), https://www.brennancenter.org/sites/default/files/2021-06/M.D.%20Ala.%2021-cv-00211%20dckt%20000116_001%20filed%202021-04-26%20Abowd%20declaration.pdf

In 2018, the Census Bureau announced its intention to move to differential privacy for the 2020 Census products (Abowd, 2018) and issued a Federal Register Notice[12] asking data users for information about what data they used, what data they needed, and what calculations they produced. To most responsibly design the system, the Census Bureau needed to know how its data were being used (Abowd & Velkoff, 2019). This request was intended to inform the implementation work. Instead, all hell broke loose.

## 3. Disclosure Avoidance Controversy

When the Census Bureau announced its intention to modernize its statistical disclosure avoidance procedures using differential privacy and invited feedback about data uses, government scientists told us that they imagined that census data users would appreciate greater transparency into the limits of the data with which they were working and relish the opportunity to provide feedback. While some users applauded the transparency, others did not. Many data users did not know what to make of this information or the request for feedback. They were uncertain of what the Census Bureau was asking for, let alone why. A broad swath of stakeholders expressed disbelief that the bureau intended to alter the data it collected; many of these stakeholders appeared unaware that the bureau had been intentionally altering the data for disclosure avoidance purposes since 1990.

For government scientists who worked on disclosure avoidance, the shift to differential privacy did not require an epistemic shift. They had been using noise infusion for 30 years. Moreover, many government scientists regularly grappled with limitations, error, and uncertainty to improve the statistical work of the agency. To these experts, the move to differential privacy was simply an operational shift that would require significant technical investment. Such an investment would, in their eyes, ensure more responsible statistical work that comported with their epistemic view. But this was not how this transition was understood externally, or even among Census Bureau staff who had never previously considered how the work of the disclosure avoidance team impacted the data. To these stakeholders, uncertainty was something to minimize, not something to strategically harness.

The announcement of the move to differential privacy, along with the stilted communications from the government, prompted a range of census data users and stakeholders to interpret the disclosure avoidance system as an existential threat. Early hair-raising proclamations by outraged stakeholders – including scholarly commentaries (Ruggles et al., 2019) and an early letter sent to the bureau, signed by hundreds of data users[13] – focused on how the Census Bureau had no right to take the public's data away. That first Federal Register Notice asked data users to articulate which data they most needed. Data users responded with force[14]: they needed *all* of the data.

---

[12] https://www.federalregister.gov/documents/2018/07/19/2018-15458/soliciting-feedback-from-users-on-2020-census-data-products

[13] This letter was sent by postal mail to multiple executives at the Census Bureau, two of whom showed the letter to one author.

Census Bureau leadership did not know what to make of this feedback or the negative backlash. The disclosure avoidance team was looking for information that would help them govern the system; they were not prepared for the very purpose of the system to be questioned. The bureau's scientists mistakenly presumed that data users would better understand and appreciate the reason for the Notice if they understood differential privacy. They believed that publishing code[15] based on differential privacy would inform and empower users.

Some technical stakeholders tried to develop ways of working with this code, but others viewed the bureau's offering as imperceptive and unhelpful. Most data users did not want code; they wanted demonstration data that could help them understand what this system would do to the data that they understood. The government responded to this request by publishing demonstration data products.[16] The scientific team viewed these as representative of a system that was a 'work-in-progress.' After the first set of demonstration data were released, the bureau asked the Committee on National Statistics to convene stakeholders evaluating the data products to share lessons.[17] At the December 2019 gathering, presenters revealed their findings, highlighting a range of issues that left them concerned. Through this process, the bureau was able to identify problems in the system[18] while also learning from data users what they wanted and needed. Yet, most external participants did not see this exercise as productive; many left that event voicing anger, frustration, and distrust, dozens of whom expressed such sentiment to one author directly and in group meetings we attended. In commentaries afterwards, the co-chairs of the event – V. Joseph Hotz and Joseph Salvo – highlighted concerns about both the system and the process (Hotz & Salvo, 2020). They summarized the main data quality issues raised by participants, including usability of data for small geographic areas and certain population groups, temporal consistency of population counts, bias introduced by post-processing, and implications for other Census Bureau projects. Noting that many community members "remain[ed] skeptical of the Bureau's adoption of differential privacy and its consequences for their use cases," Hotz and Salvo encouraged the Bureau to provide better education and guidance for data users and to "ensure meaningful involvement and feedback from the user community" (Hotz & Salvo, 2020).

In response, the Census Bureau asked both of its Federal Advisory Committees to create a working group on disclosure avoidance.[19] The bureau commissioned JASON, a group of technical advisors to the government, to evaluate the government's initiative (JASON, 2020). Non-technical stakeholders requested other means of engaging in the process: members of the National Congress of American Indians (NCAI) and the Alaska Federation of Natives (AFN) asked the Census Bureau to engage on disclosure avoidance as part of tribal consultations[20], and a range of stakeholders – including groups of civil rights advocates, mayors and governors, academics, and other federal agencies – asked the bureau to give presentations and answer questions. Census Bureau staff also began producing regular public written communications via blogs[21] and newsletters,[22] presenting at public conferences, and hosting webinars[23] on key topics and developments.

---

[14] https://www.regulations.gov/docket/USBC-2018-0009
[15] https://github.com/uscensusbureau/census2020-das-e2e
[16] The first demonstration data were released in October 2019. All demonstration data can be found online: https://www.census.gov/programs-surveys/decennial-census/decade/2020/planning-management/process/disclosure-avoidance/2020-das-development.html
[17] Details of the meeting, including the proceedings, can be found on CNSTAT's website: https://www.nationalacademies.org/our-work/2020-census-data-products-a-workshop
[18] https://www.census.gov/newsroom/blogs/research-matters/2020/03/modernizing_disclosu.html
[19] https://www.census.gov/about/cac/sac/wg-2020-data-products.html
[20] https://www.ncai.org/policy-research-center/research-data/research-recommendations

Far from assuaging critics, both the ongoing demonstration products and the varied stakeholder engagement efforts increased the animosity between many external stakeholders and the bureau. In both public and private meetings, stakeholders repeatedly accused the Census Bureau of poor communication, ineffective and untimely stakeholder engagement, and unwillingness to listen to feedback. Some of those who were frustrated with the bureau disagreed with the premise of modernizing the disclosure avoidance system (Ruggles & Van Riper, 2021), rendering any discussion of the technical matters impossible. One of us witnessed multiple heated conversations that devolved to include accusations of malice and unsavory labeling of civil servants. In our conversations and interviews, bureau leadership expressed the belief that they were being as transparent and responsive as possible while many external stakeholders voiced anger that the bureau was being intentionally opaque.

## 4. Statistical Illusions and Epistemic Disconnects

While the controversy has several facets to it, many of the concerns – including those discussed in this special issue – center on whether the data produced by the Census Bureau using the new disclosure avoidance system are of acceptable quality. At first blush, concerns about the data quality appear to be a technical matter that can be measured and evaluated. Indeed, the demonstration products were designed to allow both internal teams and external stakeholders to evaluate the data for their needs; the disclosure avoidance team sought widespread feedback to guide the development of this system. But even when evaluations of the system are expressed technically, they are shaped by values, experiences, relationships, and institutional arrangements. Evaluations rely on epistemic perspective; they are shaped by how any given stakeholder knows what they know and believes what they believe. Debates about data quality *appear* to be about technical matters, but they also involve unacknowledged epistemic disagreements that undermine efforts to reconcile technical claims. After all, the evaluation of data – like the production of data – is socially constructed (Starr, 1987).

---

[21] e.g., John A. Abowd and Victoria A. Velkoff, "Census Bureau Works with Data Users to Protect 2020 Census Data Products," https://www.census.gov/newsroom/blogs/research-matters/2020/02/census_bureau_works.html; Abowd and Velkoff, "Modernizing Disclosure Avoidance: A Multipass Solution to Post-processing Error," https://www.census.gov/newsroom/blogs/research-matters/2020/06/modernizing_disclosu.html.
[22] e.g., "First Release of Post-Baseline Quality Metrics Results," https://content.govdelivery.com/accounts/USCENSUS/bulletins/28db2c3; "Postprocessing, Consistency, and the Challenge of Negative Numbers," https://content.govdelivery.com/accounts/USCENSUS/bulletins/2924168; "Census Bureau Partners with Committee on National Statistics to Produce New Data Demonstration Files," https://content.govdelivery.com/accounts/USCENSUS/bulletins/2933574; "Now Available: New DAS Demonstration Data File," https://content.govdelivery.com/accounts/USCENSUS/bulletins/295b1da; "New Privacy-Protected Microdata Files," https://content.govdelivery.com/accounts/USCENSUS/bulletins/2a1343a.
[23] See the Census Bureau Disclosure Avoidance Webinar Series: https://www.census.gov/data/academy/webinars/2021/disclosure-avoidance-series.html

Identifying divergent epistemic viewpoints reveals why the communication struggles have only increased over time – and why an increase in the quantity of communication has not diminished tensions. Statistical illusions about census data have obscured epistemic differences for decades, and now that these illusions are being called into question, communication around data quality concerns only serves to magnify epistemic gaps. To unpack this dynamic, we consider two statistical illusions around the meaning and properties of census data that, when broken, reveal deep epistemic disconnects: first, that the released data is the objective 'ground truth,' and second, that the format of the data is a neutral choice. Both illusions were broken by the introduction of differential privacy, which not only revealed existing uncertainties in the data but also created new uncertainties for the data evaluation process.

### 4.1   What constitutes ground truth?

The demonstration products published by the disclosure avoidance team at the Census Bureau had two distinct evaluator groups. Inside the Census Bureau, a range of civil servants – including experts from fields as varied as demography, geography, economics, statistics, and computer science – were tasked to evaluate the quality of the data for both external and internal purposes, including the production of downstream data products. Outside the Census Bureau, external stakeholders from academia, civil society, and government who represented an even more diverse range of disciplines, experiences, institutions, and practices were also invited to evaluate the demonstration products with an eye to their use cases.

In developing these demonstration products, the executive team at the Census Bureau decided to use the 2010 Census Edited File (CEF) as the base data to run through the disclosure avoidance system. This approach had two notable advantages. First, this would serve as an operational test for the 2020 Census since the disclosure avoidance system would need to work with the 2020 CEF. Second, this would ensure that internal teams were evaluating the new disclosure avoidance system without being influenced by the disclosure avoidance and other editing procedures used in 2010. Yet, this choice also limited the evaluative capabilities of external stakeholders. External evaluators do not have access to the 2010 CEF; that file is considered confidential data. As a result, external evaluators could only compare the demonstration products to the published data, data that had already been altered before being published after the 2010 Census.

The Census Bureau opted to release the same demonstration products internally and externally for numerous reasons, including technical simplicity and a belief that such an approach was more transparent. Leadership at the bureau knew that they were asking external stakeholders to compare apples and oranges, but they believed that any issues raised by external evaluators could be re-evaluated internally to determine the significance of the issue. External evaluators, however, told us that either they were not told or did not understand the implications of this choice. Moreover, these external data users did not see themselves as the government's debuggers. From their perspective, they were conducting evaluations for their own sake, to determine how anxious they should be about their ability to use future data for their own work.

When external stakeholders compared the demonstration products and published data, their results did not isolate the impact of the 2020 disclosure avoidance system; biases introduced by the methods used in the published 2010 Census data polluted their analyses. After the 2019 CNSTAT meeting made clear that evaluators were not accounting for the biases of the published data, the Census Bureau attempted to inform stakeholders that they were not comparing their analyses to ground truth. Having nothing else to compare the data to, this information was not well received by either data users or other census advocates.

The issue of what constitutes 'ground truth' plagued every subsequent evaluation of demonstration products. Even if external evaluators wished to grapple with the biases introduced by previous methods, they lacked the ability to do so. To infer even high-level metrics such as the national swapping rate, which would give some indication of the bias introduced by swapping in the published 2010 Census data, required reading through the lines in the bureau's technical presentations.[24]

Without another anchor point, evaluators repeatedly published findings asserting data quality issues based on comparisons to the published data (Beveridge, 2021; Hauer & Santos-Lozada, 2021; O'Hare, 2020; Ruggles et al., 2021). Scientific debates turned into communicative spectacle after lawsuits were filed. First, some external analyses were introduced as expert reports in *Alabama v. Commerce*, where they were shown to differ from the government's internal findings. After oral arguments at the district court, a group of political scientists and statisticians released a working paper arguing that the government's system had significant issues that would harm redistricting work (Kenny et al., 2021a). The news media picked up these findings, proclaiming that "Harvard researchers recommend Census not use privacy tool" (Schneider, 2021a). This paper incensed other Harvard researchers, some of whom issued a technical rebuttal highlighting how the comparison data was not ground truth (Bun et al., 2021). Another team of political scientists who worked on redistricting issued a comparable rebuttal, also emphasizing the logical fallacies of the comparison (Wang & Goldbloom-Helzner, 2021). The district court opted to sidestep the technical issues, ruling in favor of the government on procedural grounds.

Despite 'ground truth' critiques, the authors of the working paper published their findings in *Science* (Kenny et al., 2021b), outlining changes in redistricting that would arise "in practice"[25] when using the new approach versus the published 2010 data, while ignoring biases that the published 2010 data might have introduced into the redistricting process. This publication amplified a wedge between census stakeholders who disagreed on how quality, ground truth, and uncertainty should be evaluated.

---

[24] At the Census Scientific Advisory Committee meeting on May 25, 2021, Michael Hawes and Rolando Rodríguez presented on research into alternative disclosure avoidance techniques: https://www2.census.gov/about/partners/cac/sac/meetings/2021-05/presentation-research-on-alternatives-to-differential-privacy.pdf. With this information, we infer that the 2010 swapping rate was between 2-4% nationally.
[25] This paper dismisses an earlier publication that argued that the disclosure avoidance system would not harm redistricting (Cohen et al., 2021) by stating that the authors had conducted a "theoretical" analysis based on an earlier version of the system, arguing that their "empirical" analysis shows how the system would affect redistricting "in practice." However, in conducting interviews with redistricters, we learned that the practice of redistricting involves using whatever data are published.

The more that external stakeholders made comparisons to the original published files, the more frustrated government scientists and bureau leadership grew. The bureau never viewed published data as ground truth and were surprised to learn that some external stakeholders did. Government scientists tried to remedy this misconception by highlighting the data's limits, but this infuriated many of the external stakeholders we talked with. Government scientists could not understand why external stakeholders might treat the data as ground truth even when explicitly provided such information, nor why external stakeholders would defend swapping as a technique that provided acceptable data quality, as swapping was designed to homogenize data for confidentiality at the expense of statistical accuracy. Bureau scientists knew that the shift to multi-race options in 2000 combined with shifts in the distribution of people across geography by race and Hispanic-origin meant that the statistical bias introduced by swapping would only continue to grow. With this in mind, government scientists and leadership both wanted to leverage technical advances such as differential privacy to ensure that 2020 data could be as free from statistical bias and as close to the bureau's ground truth as possible while ensuring statistical confidentiality. They believed that what they were doing was to help external data users have higher quality data.

Most external stakeholders had no reason to ever consider how previous disclosure avoidance systems might have biased published data, even if earlier scholarship had shown that techniques like swapping introduced distortion into the data (Fienberg & McIntyre, 2005). They trusted the data they received because they trusted the bureau; furthermore, the data appeared usable, even if imperfect. Stakeholders from a range of backgrounds regularly expressed anger to one of us over what felt like a bait-and-switch, where the bureau was justifying its new approach by exposing the errors in the data that it had long asked them to trust. Moreover, given that data users had worked with these previous data quite effectively, many rejected any narrative that suggested that the previous data were significantly flawed. For them, the published data were the ground truth they knew and trusted. From this vantage point, data altered by the new disclosure avoidance system were too different from what they knew for them to tolerate.

Ground truth is an illusion (Porter, 2020). Even if we bound this ideal to the rules, categories, and definitions introduced by the Census Bureau, capturing that imagined ground truth is impossible. The Census Bureau knows that its data collection processes are imperfect but believes that, given the operational and confidentiality constraints, the data collected and produced through its well-studied, systematic approach are the best data possible. Thus, it treats its confidential data as ground truth. External stakeholders know that the bureau fails to count everyone and uses various methods to fill in the gaps, but they presume that all who shared data with the Census Bureau are represented as-is in the published data.

Published data have been perceived as a 'gold standard' by which all other data should be measured, in part because of an illusion that the published data are the best anyone could have produced. External data users know that the data are imperfect, but a trusting relationship between the bureau and its users prompted most to treat the bureau's data as nearly error-free. Even once the bureau began highlighting how previous data had been altered for disclosure avoidance purposes, data users continued to treat the published data as ground truth because experience had convinced them that the published data was accurate enough.

Instead of appreciating increased transparency into the limits of the data, data users' faith in the government faltered once the bureau began speaking of statistical uncertainty. Data users did not want the government to intentionally introduce uncertainty for the sake of confidentiality; they wanted access to the bureau's ground truth. As the controversy unfolded, anything less became unacceptable.

*4.2   Appearances matter*

Just as ground truth is socially constructed, so too is the format of census data. Census data are presented as the count of individual people, even though some of those people are produced through court-approved scientific procedures designed to minimize undercounts known as 'imputation'.[26] There is no technical need for an imputation algorithm to produce individual records rather than statistical distributions; rather, this is a legal and political requirement that stems from the constitutional language of an "actual enumeration."[27]  In *Utah v. Evans,* the Supreme Court ruled that imputation constituted an enumeration so long as it was not based on statistical sampling,[28] which was prohibited per 13 U.S.C. §195.

One of the first decisions that the Census Bureau made when it began pursuing a new disclosure avoidance system was that the output of such a system would also need to produce individual records, known internally as 'microdata'. This driver of this decision was primarily operational in nature – the Census Bureau did not have the resources or time to update every technical system that used the output of the disclosure avoidance system, and those other systems required data to exist in a specific format. Yet, the decision also helped assuage a communication concern by ensuring 'semantic plausibility' (JASON, 2022). If the Census Bureau were to publish datasets that that included fractional or negative population counts, this was bound to confuse and upset external stakeholders, including lawyers and judges.

This microdata constraint was introduced before the disclosure avoidance system was built. Those building the system did not imagine that this constraint would pose serious technical hurdles, but they quickly learned otherwise. From the moment that the first demonstration product was released, some evaluators began noting that the 'post-processing' step necessary to ensure the production of microdata introduced biases into the data. Population counts of sparse geographies were more likely to become negative after a significant amount of noise was introduced; early post-processing efforts remedied this in ways that disproportionately increased the population size of small-area geographies.

Most of the quality concerns raised by those evaluating the demonstration products stemmed from this post-processing step. Inside the Census Bureau, the technical work that took place in 2020 and 2021 almost exclusively focused on reducing the harm of post-processing (JASON, 2022). Yet, as differential privacy experts explained to us, some bias was inevitable given the bureau's various constraints (e.g., microdata, block-level publication, consistency, invariants, etc.). To reduce the amount of noise introduced and minimize the impact of post-processing, the Census Bureau instead opted to select a privacy-loss budget ('epsilon') that exceeded most stakeholders' expectations.

---

[26] See, e.g., https://www.census.gov/newsroom/blogs/random-samplings/2021/04/imputation-when-households-or-group-quarters-dont-respond.html

[27] Art. I, § 2, cl. 3 of the U.S. Constitution states that apportionment shall proceed according to states' "respective numbers" as revealed by an "actual enumeration" directed by Congress.

[28] In *Utah v. Evans*, the Supreme Court gave the following justification: "[W]e need not decide here the particular limits foreseen by the Census Clause. We need only say in this instance, where all efforts have been made to reach every household, where the methods used consist not of statistical sampling but of inference, where that inference involves a tiny percent of the population, where the alternative is to make a far less accurate assessment of the population, and where consequently manipulation of the method is highly unlikely, those limits are not exceeded." *Utah v. Evans*, 536 U.S. at 479 (majority opinion).

When the bureau announced the privacy-loss budget it would use for the first data product, critics of differential privacy leveraged the prior work of differential privacy experts to lambast the bureau. For example, a paid expert witness in *Alabama v. Commerce* turned to the news media to argue that "The inventors of differential privacy regard such a high epsilon as pointless" (Schneider, 2021b). Stunned and frustrated by the politicization of their work, differential privacy experts wanted to conduct their own technical evaluation to determine how the disclosure avoidance system impacted both confidentiality and data quality.

To evaluate the system, differential privacy experts knew that they would need to better understand how the post-processing step affected both data quality and privacy. They requested the 'noisy measurements' files that revealed the noise-infused data before post-processing was introduced (Dwork et al., 2021). These computer scientists were comfortable with the fact that these files would include negative and fractional numbers, but some lawyers and redistricting experts worried that having this data out in the public would complicate the redistricting process. Redistricting, after all, is shaped by the notion that there is only one 'true' data set.[29] Having a second dataset – especially one that included negative and fractional numbers – triggered fear in the hearts of multiple lawyers we talked with.

Even as lawyers and computer scientists debated the merits and drawbacks of releasing the noisy measurements, the Census Bureau faced a different problem. In order to minimize post-processing issues, the bureau computed additional 'detailed queries' for internal use within the system. These detailed queries did not necessarily correspond to the publicly available statistics and were not meant to be released themselves; their job was to make the publicly available statistics more accurate and simpler to produce. Unfortunately, the use of these detailed queries complicated any efforts to publicize the noisy measurements.[30] Releasing noisy measurements without the detailed queries would make little sense, because users would not be able to reproduce the official, post-processed statistics from just the noisy measurements files. However, releasing the detailed queries might pose additional disclosure risk, as they could reveal sensitive information about population characteristics that were not meant to be publicly available.[31]

---

[29] *Reynolds v. Sims* (1964) mandated a "one-person, one-vote" principle which requires voting districts to be roughly equal in population. Some states have gone even further in their own legislative requirements. According to the JASON report, "[t]he concept that districts could be exactly equal in size has always ignored the intrinsic error in census counts" (JASON, 2022, p.7).
[30] See Section 5.3 of the JASON report (2022) for a detailed discussion of these issues.
[31] Note that the detailed queries are run under differential privacy constraints (and accordingly, consume some of the privacy loss budget), so they do not pose additional disclosure risk for the individual microdata. However, they could provide additional information to an adversary conducting an attack or reveal sensitive population-level information, such as the racial composition of a particular group quarters facility (JASON, 2022).

Without access to the noisy measurements, differential privacy experts were left in a bind; they knew that the noise-injection procedures guided by differential privacy were sound, but they could not guarantee that post-processing had not undermined data quality in unexpected ways. Nor were they able to assess the level of privacy protections with the selected privacy-loss budget.

In the end, differential privacy experts could not fully endorse the new system, as they were unwilling to make claims without conducting a technical evaluation and unable to get the information necessary to do so.[32] Lacking the green light from these experts, and mindful of the importance of high-quality data for minority communities, the broader community of privacy advocates also largely abstained from publicly defending the system.[33] JASON, the technical advisors commissioned by the government, highlighted how the complexity of the system and the struggles external evaluators faced inadvertently undermined the transparency the bureau was hoping to enable (JASON, 2022).

The early microdata decision by the Census Bureau may have ensured a data format that was legible to many census stakeholders, but it also complicated the technical work of producing data. Furthermore, it undermined the evaluation abilities of those who might have defended the Census Bureau's implementation.

### 4.3 Undoing statistical illusions, grappling with uncertainty

By unpacking the ways in which differential privacy clashes with the statistical illusions of ground truth and neutral data formats, we start to understand the uproar around modernizing disclosure avoidance techniques. Differential privacy is designed to add statistical uncertainty to protect confidentiality while maintaining statistical quality. This approach presumes that everyone sees the census data products that come after the count used for apportionment as a statistical product rather than a pure enumeration.[34] The epistemic struggles over ground truth and the bureau's post-processing work reveal that this is not the case. Moreover, the bureau's struggle to build legitimacy through its demonstration products and engagement work highlights how the bureau misunderstood the mechanisms that stakeholders use to build confidence in census data.

By subverting statistical illusions of objective ground truth and neutral data formats, the development of the new disclosure avoidance system both revealed and created uncertainties within the data evaluation process.  In effect, differential privacy did not just inject noise into the technical system; it also introduced uncertainty into the legitimacy-making work of the bureau.

---

[32] A group of technical privacy experts did submit an amicus brief in *Alabama v. Commerce* detailing the threats of reconstruction and reidentification attacks that highlighted why differential privacy was the correct response to such threats, thereby supporting the government's investment in such work. They did not, however, defend the Census Bureau's implementation. https://www.brennancenter.org/sites/default/files/2021-04/Amicus%20Brief_dataprivacyexperts_%202021-04-23.pdf

[33] The Electronic Privacy Information Center (EPIC) submitted an amicus brief in support of differential privacy in *Alabama v. Commerce*: https://www.brennancenter.org/sites/default/files/2021-04/Amicus%20Brief_EPIC_%202021-04-26.pdf. However, other privacy advocacy organizations remained silent on the issue.

[34] The Constitution requires that data used to determine states' allocations of representatives to Congress must be an enumeration. Supreme Court decisions – like *Department of Commerce v. U.S. House of Representatives* and *Utah v. Evans* – have clarified what constitutes an enumeration. Requirements for apportionment do not apply to other census data products.

Scientific advances and technical developments involve working with, measuring, and struggling to limit uncertainty (Beck & Wehling, 2013; Daipha, 2015; Pietruska, 2017), including technical uncertainty and perceptual uncertainty. Eliminating uncertainty is not a necessary precursor for closure; scientific consensus typically emerges when scientists understand the contours of uncertainty, not when they have eliminated uncertainty. Yet, one way to politicize a scientific consensus is to weaponize uncertainty (Kreps & Kriner, 2020). This is how corporations and politicians have, for example, managed to turn scientific consensus about the causes of climate change into a controversy (Edwards, 2013; Oreskes & Conway, 2011; Proctor & Schiebinger, 2008). The bureau's leadership viewed transparency as the antidote to the weaponization of uncertainty, but as this controversy unfolded, some inside the government began to wonder if transparency itself was the weapon. This sentiment echoed a remark in an earlier era by a Census Bureau official who noted, "Never before have I been part of an organization that produces the whips with which it gets flogged. It's masochism"  (Bryant & Dunn, 1995, p. 4).

Even as some stakeholders questioned the premise of modernizing disclosure avoidance techniques, all who wished to evaluate the system on their own terms lacked the necessary information to do so. Those accustomed to comparing data were limited to using previously distorted data as their ground truth. Those who knew how to evaluate differential privacy could not determine the impact of noise addition because of the post-processing procedures. In the end, external stakeholders grew frustrated with the bureau, and the bureau did not know how to address the technical concerns, let alone communicate in a manner that resolved the epistemic and political disconnects.

The inability for external stakeholders to evaluate the system on their own terms created an opening. Those who wished to lambast the system or undermine the bureau could use experts' uncertainty to their rhetorical advantage. Ultimately, the conflation of technical uncertainty with political and legal spin only escalated the controversy.

## 5.  The Legitimacy Work of a Statistical Imaginary

The illusions undone by the new disclosure avoidance system shed light on a larger statistical imaginary surrounding the census, whereby census data are viewed as seemingly objective and neutral. In invoking the term  *'statistical imaginary'*, we are referring to and building on Sheila Jasanoff and Sang-Hyun Kim's notion of *'sociotechnical imaginaries'*, which they define as "collectively held, institutionally stabilized, and publicly performed visions of desirable futures, animated by shared understandings of forms of social life and social order attainable through, and supportive of, advances in science and technology" (Jasanoff & Kim, 2015, p. 4). Statistics are one such advance in science and technology. Like sociotechnical imaginaries, statistical imaginaries are collectively produced, standardized, and performed visions of the future. Yet in expanding on this concept, we are also seeking to capture how statistics are actively co-constructed alongside certain sociotechnical imaginaries. We introduce the term 'statistical imaginaries' to emphasize how some sociotechnical imaginaries are entangled with the ways in which the state – through statistics – has the authority to 'see' and govern its subjects (Daston & Galison, 2010; Porter, 2020; Scott, 1999). The performances around statistics in both governmental and non-governmental contexts reinforce the legitimacy of a state. Because of these entanglements, when statistical imaginaries start to unravel, so too does the legitimacy of the state.

In the U.S., the tight coupling of state and statistics can be traced back to the Constitution itself. In Article 1, Section 2, the Constitution designates the census as democracy's data infrastructure, rendering it responsible for the allocation of resources and power. In doing so, the Constitution created a statistical imaginary around the census, outlining a vision of a fair republic rooted in statistical data. The centrality of the census in the political configuration of the United States ensures that census data will be a site of political contestation; the statistical imaginary surrounding this configuration ensures that political contestations can easily manifest as battles over statistics. In other words, controversies over the statistical work of the state can – and are likely to – appear technical even when they are much more. Political and legal fights over census data have repeatedly borne this out (Bouk & boyd, 2021).

To resist the politicization that this configuration makes possible, government scientists and the statistical community have doubled down on mathematical rigor while focusing on professionalization and transparency as tools of legitimacy-making. The government encourages public trust in state-sponsored numbers through these practices and procedures (Porter, 2020), even if such performances require the maintenance of statistical illusions. The statistical imaginary that surrounds census data emerged to reinforce the state's legitimacy. Yet, this statistical imaginary is also precarious, dependent on statistical illusions, and easily shattered by those wishing to undermine the legitimacy of the state.

Science and Technology Studies (STS) scholars have long called into question the fetishization and performance of statistics by the state (Daston & Galison, 2010; Hacking, 1990; Porter, 2020; Scott, 1999). This lineage of scholarship has focused on revealing illusions, not to undermine the state but to strengthen it. Yet, this scholarship has not fully examined how undoing these illusions – without strategically managing the stabilizing work of statistical imaginaries – might harm the legitimacy of the state. Other scholarly lineages have focused on how destabilizing knowledge claims can serve political agendas (Michaels & Monforton, 2005; Proctor & Schiebinger, 2008) and how transparency can be weaponized (Levy & Johns, 2016; Pozen, 2018). Taken together, these bodies of work reveal how surfacing epistemic contestations can accelerate projects of undoing, including the undoing of the state.

Those close to census data have struggled with the illusions and imaginaries surrounding the census throughout history. Critics have long questioned the quality and fairness of census data – and the Census Bureau's objectivity and neutrality (M. J. Anderson & Fienberg, 2001; Nobles, 2000; Schor, 2020). Debates about undercounts date back over a century (Bouk & boyd, 2021). Yet, rarely do these challenges call into question the role of statistics because the 'fix' offered by both critics and the bureau is to improve data collection procedures (M. J. Anderson, 2015; O'Hare, 2019; Rosenthal, 2000) and compensate for limitations of the data (Groves & Lyberg, 2010). Most census experts are not seeking to undermine the legitimacy of the state; they are invested in increasing the quality and fairness of the data.

In other cases, however, politicians have waged legal and political fights over statistical practices, not simply to improve data quality. Partisan actors have also used technical disagreements for strategic purposes. For example, statisticians in the 1920s thought they were debating method to ensure neutrality and fairness, even as their fight was leveraged by politicians driving xenophobic agendas (Bouk, 2021; Bouk & boyd, 2021). Scientific disagreements over sampling and statistical adjustment turned into legal and Congressional battles purportedly about statistical methods, even though they were really about political control (M. J. Anderson & Fienberg, 1999; Brown et al., 1999; Choldin, 1994). In each of these fights, there were significant technical and scientific disagreements, but spectacle was formed because of political and partisan agendas.

In some ways, the controversy surrounding differential privacy parallels earlier fights about statistical methods, including the so-called 'adjustment wars,' where both technical disagreements and political calculations were at play. Even stakeholders who share the same ideological values founds themselves at odds when such contestations emerge. Yet, what makes this particular controversy more complicated stems from the far-reaching nature of this epistemic rupture, triggered not only by the proposal of differential privacy but also by its implementation. The change in procedure has taken place, demanding an epistemic transition that many are not ready to make. Divergent mutually exclusive epistemic frameworks cannot sustain a statistical imaginary that upholds the legitimacy of the census.

Differential privacy is designed to balance statistical accuracy and statistical confidentiality. It is predicated on an epistemology of probabilistic models and statistical claims, one in which uncertainty is a given. Those who embrace differential privacy acknowledge that data are already imperfect; they treat uncertainty as something that can be productively harnessed to balance competing moral, technical, and legal commitments and encourage technical modes of governance, such as setting parameters to tune the tradeoff between confidentiality and detail-level accuracy. This explicit acknowledgement that data must be governed undoes illusions of objectivity and neutrality, calling into question the collectively maintained statistical imaginary surrounding the census. Those who embrace differential privacy believe that these statistical illusions are harmful, as they hinder the evaluation and improvement of data quality and are at odds with an epistemic framework that centers uncertainty. Those who reject this approach view centering uncertainty as politically, legally, and rhetorically dangerous; they prefer statistical illusions, however technically imperfect. Unfortunately, these divergent perspectives are not compatible. Failure to make this transition increases data quality issues but embracing uncertainty requires significant political upheaval. This no-win situation ensures epistemic fractures that can easily be politicized.

By implementing differential privacy without first shifting the statistical imaginary to account for uncertainty, the Census Bureau has created a flashpoint, whether or not it intended to. Many census stakeholders – both inside the government and external to it – are struggling with the implications of this transition. While the communication and technical issues are vast, they stem in part from the bureau's attempt to simultaneously adhere to the existing statistical imaginary while also disrupting it. The technical issues around post-processing reflect the bureau's desire to deploy transparent techniques to protect statistical confidentiality while maintaining the pretense that a modern census is simply a count. In its attempt to salvage this statistical illusion, the bureau both created additional technical problems and hindered analyses that may have helped legitimize the work. The communications issues around what constitutes ground truth also reflect how the bureau's approach to creating statistics is not aligned with external illusions about the making of data.

Most census stakeholders – including government officials and census advocates – believe in transparency. But how individual stakeholder groups understand transparency flows from how they understand the construction of knowledge. Transparency by itself is meaningless; it can only strengthen trust when all involved are working within a mutually understandable epistemic framework. The bureau's technical transparency was sensible to computer scientists but made little sense to most external stakeholders, who interpreted the bureau's inability to communicate within non-statistical epistemic frameworks as mal-intentioned. The more frustrated and resentful each side grew with the other, the more fractured the epistemic landscape became – and the easier it was for those who sought to take advantage of this controversy to destabilize the legitimacy of the census.

A reckoning over the census' statistical imaginary predicated on objectivity and neutrality was likely to occur eventually; the pressure for more data and the procedural, legal, moral, and technical need for statistical confidentiality have been on a collision course for decades. The controversy over differential privacy during a period of broader institutional and political instability has simply hastened that reckoning. Censuses have certainly weathered controversies larger than the one around differential privacy, but this epistemic fight has emerged at a time when our democracy is under attack from all directions. The 2020 Census was rife with a range of other controversies and challenges, including the Supreme Court case about including a citizenship question,[35] political interference during the Trump Administration,[36] lawsuits over the schedule,[37] and the Covid-19 pandemic.[38] Each of these chiseled away at the legitimacy of the Census Bureau and the 2020 Census data. These issues have come to an entangled head around the 2020 Census's disclosure avoidance system, which makes repairing the ruptures of this controversy all the more challenging – yet crucial – for the legitimacy of future censuses and the institution producing them.

As with all controversies, there are different views on the best path to closure. Some see the path forward with a nostalgic eye, seeking to return to the illusions of the past. Others see an opening to minimize the data work of federal statistical agencies while others see an opportunity to reconsider privacy provisions and increase data-sharing by statistical agencies. Still others wish for uncertainty to be embraced as a cornerstone of the government's legitimacy. None of these paths forward can be seen exclusively in technical terms, for they are all epistemic strategies to pursue or thwart legitimacy.

---

[35] *Department of Commerce v. New York*, 2019
[36] Michael Wines, "Census Memo Cites 'Unprecedented' Meddling by Trump Administration," https://www.nytimes.com/2022/01/15/us/2020-census-trump.html
[37] *State of Ohio v. Department of Commerce*, 2021: https://www.ohioattorneygeneral.gov/Files/Briefing-Room/News-Releases/Activity-in-Case-321-cv-00064-TMR-State-Of-Ohio-v.aspx; *Alabama v. Department of Commerce*, 2021: https://www.brennancenter.org/our-work/court-cases/alabama-v-united-states-department-commerce
[38] Michael Wines, "Census Suspends Field Operations Amid Coronavirus Fears," https://www.nytimes.com/2020/03/18/us/virus-census-homeless.html

There are also plenty who have chosen to stay out of the fray entirely. At the end of the day, most census data users are not engaged directly with this controversy, although the ephemeral legitimacy battles continue to influence their perception of the data. These data users simply want to have and use census data for a range of important purposes. To that end, they have one simple question: is 2020 Census data good enough for them to use? As this controversy unfolds and the statistical imaginary becomes increasingly brittle, more and more data users are being forced to reckon with technical questions about statistical quality and epistemic questions about how to know and trust the data they are given. They too are being asked to grapple with how they know what they know. Yet, for many, this is unfamiliar territory.

Unfortunately, true closure to this controversy is unlikely to come from technical discussions, better communication, or legal battles. After all, fights about differential privacy are not solely about technical matters; they involve conflicting epistemic notions of how census data are produced, how technical uncertainty is managed, and what the state can govern. Fundamentally, they are struggles over what the statistical imaginary around the census should be. Transparency alone does not bridge these gaps; it often magnifies them. Legal battles and policy interventions, too, will not resolve the present entanglements, let alone generate trust in the Census Bureau. Over time, data users may come to accept that data governed by differential privacy are just as usable as previous data. Yet, it is also possible that this controversy will continue to escalate, serving as a wedge to the legitimacy of the census, and undermining trust in the federal government and its statistical infrastructure.

Moving forward constructively requires an epistemic reckoning, one that involves shifting the statistical imaginary from seeing data as neutral and objective to recognizing that data will always have limitations and must be governed. Regardless of how disclosure avoidance is managed, governing data means all involved must collectively grapple with uncertainty, contend with what makes data legitimate and trustworthy, and deploy new tools to resist the politicization of data and statistics. Such a shift will not arise simply through technical or organizational transformations. Only by remaking our statistical imaginary into one that is grounded in a responsible approach to data can we hope to repair the legitimacy of our democracy's data infrastructure.

**Acknowledgements**

This paper was made possible thanks to countless conversations with advocates, demographers, statisticians, and Census Bureau employees who patiently tried to explain their perspectives, their work, and why they felt like others didn't understand their concerns. We are indebted to the anonymous reviewers and HDSR editors who took the time to read drafts of this paper and critically engage with the arguments. We are especially grateful to those who reviewed drafts of this paper and provided detailed comments, including John Abowd, Dan Bouk, Connie Citro, Damien Desfontaines, Abie Flaxman, Matt Goerzen, Danny Goroff, Brent Hecht, Jessica Hullman, Ron Jarmin, Steve Jost, Sallie Keller, Gary King, Jae June Lee, Margaret Levenstein, Muira McCammon, Jake Metcalfe, Terry Ao Minnis, Manny Moss, Kathy Pettit, Ken Prewitt, Rida Qadri, Burton Reist, Alex Rosenblat, Seth Spielman, Ryan Steed, Rebecca Tippett, Fred

Turner, Sam Wang, Moira Weigel, and Tom Wolf. Research assistants make the scholarship go round; Kevin Ackermann has played a pivotal role in making this paper possible.

The bulk of this work was funded by our home institutions: Microsoft Research and Harvard University. While this paper is our original scholarship, a significant amount of data collection and thinking was made possible through boyd's active participation in two funded collaborative projects. The first, conducted alongside Dan Bouk, focused on historical case studies of the census; this was supported by the Alfred P. Sloan Foundation (G-2019-12414). The second was done in conjunction with the Disinformation Action Lab (Charley Johnson, Cristina López G., Emma Margolin, William Partin) at Data & Society (which boyd founded and currently serves as president of the board of directors). The initial gifts for this effort were provided by the John S. and James L. Knight Foundation and Arabella Advisors. Data & Society's full list of funders can be found online at https://datasociety.net/wp-content/uploads/2021/01/Funders-List-2021.pdf. Support for Kevin Ackermann, who served as the research assistant supporting this work, also comes from Data & Society.

The U.S. Census Bureau supported this ethnographic study by giving danah boyd access to observe the government's work during the 2020 Census, compliant with Special Sworn Status. Her study was certified by Microsoft Research's Institutional Review Board (IORG #0008066, IRB00009672, #577/630). Data & Society study was certified by Advarra's IRB (IORG #0000635, IRB00000971, #10-184). Jayshree Sarathy is partially supported by Cooperative Agreement CB20ADR0160001 with the Census Bureau. The views expressed in this paper are those of the authors and not those of the U.S. Census Bureau or any other sponsor.